\documentclass[conference,final]{IEEEtran}

\usepackage{adjustbox}
\usepackage{amsfonts}
\usepackage{amssymb}
\usepackage{amsthm}
\usepackage{mathrsfs}
\usepackage{xfrac}
\usepackage{pgf}
\usepackage{tikz}
\usetikzlibrary{arrows,automata,positioning,calc}
\usepackage{graphicx}

\usepackage{thm-restate}

\usepackage[margins]{trackchanges}
\addeditor{FS}
\addeditor{LuS}
\addeditor{LiS}
\addeditor{AC}
\newcommand{\fs}[1]{\note[FS]{#1}}

\makeatletter
\def\thm@space@setup{%
  \thm@preskip=10pt \thm@postskip=\thm@preskip
}
\makeatother

\newtheorem{definition}{Definition}
\newtheorem{theorem}{Theorem}
\newtheorem{example}{Example}
\newtheorem{corollary}{Corollary}
\newtheorem{proposition}{Proposition}

\newcommand{\MDPstates}{S}
\newcommand{\MDPact}{\textit{Act}}
\newcommand{\MDPprob}{{Pr}}
\newcommand{\MDPinit}{\iota}
\newcommand{\MDPlabel}{L}
\newcommand{\MDP}{(\MDPstates, \MDPact, \MDPprob, \MDPinit, \MDPlabel)}
\newcommand{\MDPi}[1]{(\MDPstates_{#1}, \allowbreak \MDPact_{#1}, \allowbreak \MDPprob_{#1}, \allowbreak \MDPinit_{#1}, \allowbreak \MDPlabel_{#1})}
\newcommand{\proc}{\mathcal{M}}
\newcommand{\procslice}{\mathcal{M}^{\textit{slice}}}

\newcommand{\AP}{\textit{AP}}
\newcommand{\act}{\alpha}
\newcommand{\Prob}{\mathbb{P}}
\newcommand{\Probmin}{\mathcal{P}_{min}}
\newcommand{\Probmax}{\mathcal{P}_{max}}

\newcommand{\true}{\textit{true}}

\newcommand{\equiprob}{\epsilon}
\newcommand{\quotient}[2]{\sfrac{#1}{#2}}

\newcommand{\PCTLstar}{\textit{PCTL}^\star}

\newcommand{\ctlnext}{X}
\newcommand{\ctlglob}{G}
\newcommand{\ctlfuture}{F}

\newcommand{\var}[1]{\textsf{#1}}
\newcommand{\varval}[1]{\textsf{#1}}
\newcommand{\prop}[1]{\textsc{#1}}

\newcommand{\Nat}{\mathbb{N}}
\newcommand{\NatPos}{\Nat_{> 0}}

\newcommand{\bsim}{\approx}

\newcommand{\pre}{\textit{pre}}
\newcommand{\post}{\textit{post}}

\newcommand{\tguard}{\textit{guard}}
\newcommand{\Client}{\textsc{Client}}

\newcommand{\SliceAttacker}{\textsc{SliceAtt}}
\newcommand{\ServerAttacker}{\textsc{ProviderAtt}}

\newcommand{\comp}{\parallel}

\newcommand{\newcolumn}{\vfill\eject}

\newif\ifextended
\extendedtrue

\title{Parametric and Probabilistic Model Checking of Confidentiality in Data Dispersal Algorithms\ifextended\\(Extended Version)\fi}

\author{
    \IEEEauthorblockN{Marco Baldi, Alessandro Cucchiarelli, Linda Senigagliesi, Luca Spalazzi, Francesco Spegni} \\
    
    \IEEEauthorblockA{Dipartimento di Ingegneria dell'Informazione \\ Universit\`a Politecnica delle Marche \\ Ancona, Italy}
}

\begin{document}

\maketitle

\begin{abstract}
Recent developments in cloud storage architectures have originated new models of online storage as cooperative storage systems and interconnected clouds. Such distributed environments involve many organizations, thus ensuring confidentiality becomes crucial: only legitimate clients should recover the information they distribute among storage nodes.

In this work we present a unified framework for verifying confidentiality of dispersal algorithms against probabilistic models of intruders. Two models of intruders are given, corresponding to different types of attackers: one aiming at intercepting as many slices of information as possible, and the other aiming at attacking the storage providers in the network. Both try to recover the original information, given the intercepted slices.

By using probabilistic model checking, we can measure the degree of confidentiality of the system exploring exhaustively all possible behaviors. 
Our experiments suggest that dispersal algorithms ensure a high degree of confidentiality against the slice intruder, no matter the number of storage providers in the system. On the contrary, they show a low level of confidentiality against the provider intruder in networks with few storage providers (e.g. interconnected cloud storage solutions).
\end{abstract}


\section{Motivation}
\label{sec:motivation}

\fs{check missing "s" for 3rd person all over the paper}

Recently, the spreading of online storage services (such as iCloud, Dropbox, Skydrive, etc.) has seen a huge increase. 
According to the classical paradigm, the service provider buys or rents a large number of servers in which authorized clients are able to store their data. Recently, two other paradigms have emerged, viz. cooperative storage services (CSS) and federated (storage) clouds (FC). \fs{find citation}

In CSS, the storage capacity is provided directly by the clients themselves who, co-operating in the cloud, make their own storage facilities available to the others. This approach offers some evident advantages: first of all the service provider only needs a small number of servers, acting as coordinators for the access to the service. Secondly, increasing the number of users yields an increase of the storage capabilities. In this context, data is stored by the users, each acting as a \emph{storage server}.
In FC, organizations decide to agree on sharing their resources for solving common tasks. In this scenario, data is read from (and written to) several \emph{storage providers}, each managing its set of storage servers behinds its cloud.

A fundamental requirement of CSS and FC is confidentiality: only the legitimate client should be able to recover the original information.

A consolidated solution to achieve confidentiality in such contexts is based on data dispersal. Dispersal algorithms provide a methodology for storing information in $n$ distinct pieces, or slices, (dispersed) across multiple locations, so that redundancy protects the information in the event of a location outage, but unauthorized access at any single location does not provide usable information. Only the originator or a user which has access to, at least, $k$ out of the $n$ slices distributed among $m$ available servers (or providers) can properly assemble and recover the complete information, without the need of any pre-shared encryption key. Instead, a client or attacker retrieving a number of slices lower than $k$ is not able to get any information. This basic principle has been applied since the pioneer works by Shamir \cite{Shamir1979} and Rabin \cite{Rabin1989}, and subsequently confirmed by McEliece and Sarwate \cite{McEliece1981} who disclosed the relationship with Reed-Solomon (RS) coding schemes. 

Dispersal algorithms based on RS schemes have optimal performances, but they are subject to constraints due to the algebraic nature of the codes which practically limit the number of servers $m$, i.e., the level of dispersion. Since a large number of servers is convenient both from the client point of view (which can tolerate a greater number of failures) and from the security point of view (since the attacker needs to crack a larger number of nodes to steal the data), new solutions based on the Luby transform (LT) codes for the dispersal algorithm have been proposed \cite{Luby2002}. These codes have no limit, in principle, on the value of $n$. On the other side, a characteristic feature of LT is that there are two thresholds $k_1$ and $k_2$ such that if an attacker has access to at least $k_1$ slices it has some probability of reconstructing the entire message, while if it has access to at least $k_2$ slices it has all the needed information to reconstruct the message. For this reason, one can see the RS coding schemes as special cases of LT.\fs{check}

Both coding schemes, RS and LT, can be combined with an all-or-nothing-transform (AONT) \cite{Rivest1997b} realising AONT-RS and AONT-LT \cite{Baldi2014} data dispersal algorithms. \fs{check}

In this work we provide a unified formal framework to model check the mentioned dispersal algorithms against different kinds of attackers trying to intercept slices and reconstruct the original message. We use the tool PRISM \cite{Prism2011} to verify the degree of confidentiality of such algorithms.
Since the problem is intrinsically parametric, we also want to identify suitable conditions under which the verification outcome holds for any number of storage providers in the cloud. To this aim, we repeatedly measured the probability of a confidentiality attack to understand how it varies w.r.t. other parameters, and in particular when the number of slices $n$ increases.

Two different types of intruder are analyzed: the first one can eavesdrop the slices traveling to a server without interfering with the communication. The second type of intruder can violate some providers and retrieve all the slices they store. Both intruders are assumed to be \emph{passive} and \emph{probabilistic}, meaning that they can only read the exchanged information, and probabilities affect their capability of taking any action.

The organization of the paper is as follows: Section \ref{sec:modeling} defines the modeling languages and the models used for our analysis; Section \ref{sec:formal} shows the formal analysis allowing us to measure the probability of an attack; Section \ref{sec:experiments} comments the experimental data; Section \ref{sec:rw} compares our work with the existing literature, while Section \ref{sec:conclusions} summarizes our results. \ifextended\else\footnote{The detailed proofs of can be found in the extended version~\cite{SPCloudExtended}.}\fi

\section{Modeling}
\label{sec:modeling}

Here, we formally describe a client process and two types of intruder. The main client responsibility is sending a sequence of \emph{slices} to several distinct storage servers, following the specific dispersal algorithm. The set of slices constitutes the \emph{message}. A \emph{message} includes an actual content, viz. \emph{message body}, and some extra information, viz. \emph{message payload}, containing, among other things, the replicated information allowing to reconstruct the message body even in presence of faulty storage nodes. The main responsibility for both types of intruder is to intercept the traveling slices, and reconstruct the message body. The key difference between them is that the first type intercepts every slice independently from the previous ones. The second type of intruder, on the other side, attacks the storage providers and collect all the stored slices. The actions of the intruders are probabilistic.

The following parameters affect the system behavior: $n$ is the number of slices that compose the message, $m$ is the number of storage providers (or servers) in the system, $c$ is the number of slices every server/provider can store, also called \emph{capacity}, $k_1$ and $k_2$ are two thresholds such that if the attacker intercepts at least $k_1$ slices it has \emph{some} probability of reconstructing the entire message body, while if it intercepts at least $k_2$ slices it has \emph{all} the necessary information to reconstruct the message body. Two series of probabilities are used: $a_i$ is the probability of intercepting a slice traveling to storage provider $i$, for the first attacker, and it is the probability of attacking the storage provider $i$, for the second attacker; $x_j$ is the probability of reconstructing the entire message body, given $j$ slices have been intercepted by the attacker. The relation among parameters are the following: $k_1 \le k_2 \le n$ and $n \le m \cdot c$. Probabilities $a_i$ are defined for $i \in [1,m]$, while probabilities $x_j$ are defined for $j \in [k_1,n]$ and is such that: $\forall j \in [k_1,k_2-1].\ 0 < x_j < 1$ and $\forall j \in [k_2,n].\ x_j = 1$. Also: $\forall j \in [k_1,n].\ x_j \leq x_{j+1}$.

\subsection{The system}

\emph{Markov Decision Processes}, or MDP for short, is a formalism allowing the definition of systems with probabilistic and non-deterministic actions. They are thus recognized as a good means to model randomized distributed systems: each process is described by its probabilistic transition function, and processes are interleaved by a non-deterministic scheduler \fs{here we take care of observation by Reviewer 3}. We briefly introduce MDPs using Baier and Katoen's notation \cite{BaierKatoen}.

\begin{definition}[\bf Markov Decision Process]
Assume a finite set of atomic propositions $\AP$. A \emph{Markov decision process} is a tuple $\proc = \MDP$ where:
\begin{itemize}
\item $\MDPstates = \{ s_1, s_2, \ldots \}$ is a finite set of states,
\item $\MDPact = \{ \act_1, \act_2, \ldots \}$ is a finite set of actions, 
\item $\MDPprob : \MDPstates \times \MDPact \times \MDPstates \to [0,1]$ is a probabilistic transition function such that $\sum_{s' \in S} \MDPprob(s,\act,s') \in \{ 0,1 \}$, for all $s \in \MDPstates$ and $\act \in \MDPact$;
\item $\MDPinit : \MDPstates \to [0,1]$ is the initial distribution probability of states, such that $\sum_{s \in \MDPstates} \MDPinit(s) = 1$; 
\item $\MDPlabel : S \to 2^\AP$ is a labeling function.
\end{itemize}
\end{definition}

We write $(s,\alpha,p,s') \in \MDPprob$ whenever $\MDPprob(s,\alpha,s') = p$, for some suitable states $s,s'$, action $\alpha$ and probability $p > 0$. We call \emph{transition} any such tuple.

If the MDP is in state $s$, an action $\act$ is \emph{enabled} if $(s,\act,p,s') \in \MDPprob$, for some state $s'$ and probability $p$. If an action $\act$ is enabled, than the probabilities among $\act$-transitions must form a probability distribution: $\sum_{s' \in S} \MDPprob(s,\act,s') = 1$. More than one action can be enabled in the same state $s$, thus the sum of probabilities of all transitions leaving state $s$ sum up to the number of enabled actions. Let us remark that while probabilities in MDP could be real values, for algorithmic purposes in this work we constrain them to be rational values.

In the following we make use of the uniform probability distribution $\equiprob_m : [1,m] \to [0,1]$ having the property: $\forall i \in [1,m].\ \equiprob_m(i) = \frac{1}{m}$. We write states of MDPs as configurations of some given set of variables $V$. Given a state $s$ and a variable $v \in V$, we write $s.v$ to denote the value of the variable in that state. Given two states $s$ and $s'$ and a set of variables $V$, we write $s \equiv_V s'$ meaning that the values of variables in $V$ are the same in both states: $\forall v \in V.\ s.v = s'.v$. Given a labeling $L : S \to 2^\AP$, we say that $L$ is \emph{invariant} w.r.t. $V$ iff $\forall s,s' \in S.\ s \equiv_V s' \Rightarrow L(s) = L(s')$.

Given MDPs $\proc_1$ and $\proc_2$, we will denote with $\proc_1 \comp \proc_2$ the MDP resulting from their \emph{synchronous composition}.

Let us list the graphical conventions used in this work to depict MDPs (see Fig. \ref{fig:client} and following):
\begin{itemize}
\item the circles represent the values of variables $\var{pc}_c$ or $\var{pc}_a$;
\item transitions have labels of the form: $\{ p \} [ \act ] \gamma$ where $p \in [0,1]$ is the probability, $\act$ is the action, and $\gamma$ is a boolean formula mentioning the variables of the state.
\end{itemize}
For the sake of brevity, we use notation $\exists i \in [a,b]. \allowbreak \{ p(i) \} [\act] \gamma(i)$ as transition label, to denote a group of $(b-a+1)$ similar transitions, each obtained by replacing $i$ with one of the natural values in the interval $[a,b]$. Following the PRISM notation, the boolean formula on a transition can refer to variables in the source state by their name (e.g. $\var{ctr}_c$), and to variables in the target state by their primed name (e.g. $\var{ctr}'_c$).

\subsection{The client}

\begin{figure}[tb]
\centering
\resizebox{0.7\linewidth}{!}{\begin{tikzpicture}[->,>=stealth',shorten >=1pt,auto,node distance=5.5cm,
                    semithick]
  \tikzstyle{every state}=[circle, minimum width=30pt, fill=white,draw=black,text=black]
  \tikzstyle{ghost}=[fill=white, draw=none,text=black]

  \node[state]             (A)                                      {$0$};
  \node[state]             (B) [above right = 2.5cm and 4cm of A]   {$1$};
  \node[state]             (C) [below right = 2.5cm and 4cm of A]   {$1$};
  \node[state]             (D) [left = 2cm of A]                    {$2$};
  \node[ghost]             (E) [above left = 3.5cm of A]            {};
  \node[ghost]             (F) [right = 4cm of A]                   {$\vdots$};
 
  \tikzstyle{lblabove}=[sloped,above,midway,align=center] 
  \tikzstyle{lblbelow}=[sloped,below,midway,align=center]

  \path (A) edge              node[lblabove] {$\{ p_1 \} ~ \var{ctr}_c < n \wedge \var{s}_c' = 1$}
                                                                (B)
            edge              node[lblabove] {$\{ p_m \} ~ \var{ctr}_c < n \wedge \var{s}_c' = m$}
                                                                (C)
            edge              node[lblabove] {$\var{ctr}_c = n$}  (D)
        (B) edge [bend left]  node[lblbelow] {$\var{ctr}_c^1 \ge c$} 
                                                                (A)
            edge [bend right] node[align=center, sloped, midway, above] {$[busy] \var{ctr}_c^1 < c \wedge$\\$\var{ctr}_c' = \var{ctr}_c+1 \wedge {\var{ctr}'}_c^1 = {\var{ctr}'}_c^1 + 1$}
                                                                (A)
        (C) edge [bend left]  node[lblbelow] {$[busy] \var{ctr}_c^m < c \wedge$\\$\var{ctr}'_c = \var{ctr}_c+1 \wedge {\var{ctr}'}_c^m = {\var{ctr}'}_c^m + 1$}
                                                                (A)
            edge [bend right] node[lblabove] {$\var{ctr}_c^m \ge c$} 
                                                                (A)
        (E) edge              node[near start,align=center] {$\forall l. \var{ctr}_c^l = 0 \wedge$\\$\var{ctr}_c=0 \wedge$\\$\var{s}_c=0$}                        
                                                                (A);
                
\end{tikzpicture}}
\caption{\label{fig:client}The MDP $\Client$}
\end{figure}
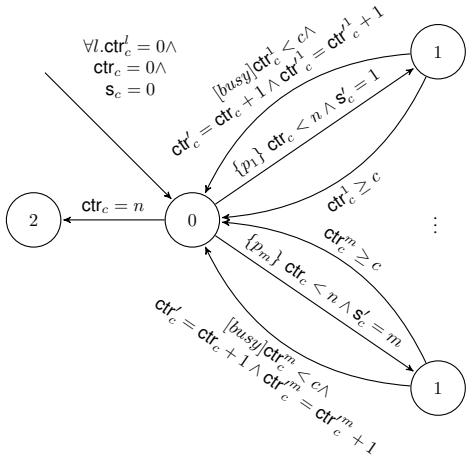

\fs{fix bugs in the figure} Let $\Client$ be the MDP encoding the client process described earlier (see Fig. \ref{fig:client}). It has the following local variables: 
\begin{itemize}
\item \makebox[0.5cm]{$\var{pc}_c$}     : track the progress of the process,
\item \makebox[0.5cm]{$\var{s}_c$}      : the identifier of the next recipient server/provider,
\item \makebox[0.5cm]{$\var{ctr}_c$}    : the total number of sent slices,
\item \makebox[0.5cm]{$\var{ctr}_c^i$}  : the number of slices sent to server/provider $i$.
\end{itemize}

The MDP has a first block of $m$ transitions from $\var{pc}_c = 0$ to $\var{pc}_c = 1$ picking a storage server (or provider) to store the slice; each transition is subject to some probability $p_i$. Next, the client either sends the slice to the selected recipient, if the latter has not reached its capacity, or it tries again picking another one. A total slice counter and a server/provider slice counter are increased whenever the slice is sent. The loop terminates when all the slices are sent.

Let us remark that the sending transition is labeled with a special action $busy$. This is used when building the synchronous composition of $\Client$ with the MDP modeling the intruder, to synchronize the action of sending by the client and the action of intercepting by the intruder.

\subsection{The slice attacker}

\begin{figure*}[tb]
\centering
\resizebox{\linewidth}{!}{\begin{tikzpicture}[->,>=stealth',shorten >=1pt,auto,node distance=5cm,
                    semithick]
  \tikzstyle{every state}=[circle, minimum width=30pt, fill=white,draw=black,text=black]
  \tikzstyle{ghost}=[circle, minimum width=30pt, fill=white,draw=white,text=black]

  \node[state]     (A)                      {0};
  \node[state]     (B) [right of=A]         {1};
  \node[state]     (C) [right=2cm of B]     {2};
  \node[state]     (D) [right of=C]         {1};
  \node[state]     (E) [right=2cm of D]     {2};
  \node[ghost]     (F) [right of=E]         {$\cdots$};
  \node[state]     (G) [above=2cm of F]     {$\var{done}$};
  \node[ghost]     (H) [left=1.5cm of A]    {};
  
  \tikzstyle{lblbelow}=[midway,below,align=center]

  \path (A) edge node[lblbelow] {$\exists i \in [1,m]. \{ a_i \} [busy]$\\$\var{s}_c = i \wedge \var{ctr}_a = k_1 - 1 \wedge$\\$\var{ctr}'_a = k_1$}
                                                                (B)
            edge [loop above, align=center, midway, above] node {$\exists i \in [1,m] . \{ a_i \}[busy]$\\$\var{s}_c=i \wedge \var{ctr}_a < k_1 - 1 \wedge$\\$\var{ctr}_a = \var{ctr}'_a + 1$}
                                                                (A)
            edge [loop below, midway, below, align=center] node {$\exists i \in [1,m].$\\$\{ 1-a_i \}[busy]$\\$\var{s}_c = i \wedge \var{ctr}_a < k_1$}                        
                                                                (A)
        (B) edge node[lblbelow] {$\{ 1 - x_{k_1} \}$}
                                                                (C)
            edge [bend left=20]  node[near start,above,align=center] {$\{ x_{k_1} \}$}
                                                                (G)
        (C) edge              node[lblbelow] {$\exists i \in [1,m]. \{ a_i \} [busy]$\\$\var{s}_c=i \wedge \var{ctr}_a = k_1 \wedge$\\$\var{ctr}'_a = k_1 + 1$}
                                                                (D)
            edge [loop below] node[lblbelow] {$\exists i \in [1,m].$\\$\{ 1-a_i \}[busy]$\\$\var{s}_c=i \wedge \var{ctr}_a = k_1$}
                                                                (C)
        (D) edge              node[lblbelow] {$\{ 1 - x_{k_1 + 1}\}$}
                                                                (E)
            edge [bend left=20]  node[near start] {$\{ x_{k_1 + 1} \}$}
                                                                (G)
        (E) edge              node[lblbelow] {$\exists i \in [1,m]. \{ a_i \} [busy]$\\$\var{s}_c=i \wedge \var{ctr}_a = k_1 + 1 \wedge$\\$\var{ctr}'_a = k_1 + 2$}
                                                                (F)
            edge [loop below] node[lblbelow] {$\exists i \in [1,m].$\\$\{1 - a_i\}[busy]$\\$\var{s}_c=i \wedge \var{ctr}_a = k_1 + 1$}
                                                                (E)
        (F) edge              node[midway,align=center] {$\exists i \in [1,m]. \{ a_i \} [busy]$\\$\var{s}_c=i \wedge \var{ctr}_a = k_2 - 1 \wedge$\\$\var{ctr}'_a = k_2$}
                                                                (G)
        (H) edge              node {$\var{ctr}_a = 0$}          (A);
				
\end{tikzpicture}
\fs{make the values of pc explicit in this figure}}
\caption{\label{fig:attacker1}The MDP $\SliceAttacker$}
\end{figure*}

Let us name $\SliceAttacker$ the MDP encoding the first type of intruder. The reason for its name is that it tries to intercept every slice, independently from the previously intercepted ones. The intruder is given in Fig. \ref{fig:attacker1}. It has two local variables, viz. $\var{pc}_a$ and $\var{ctr}_a$. The former tracks the progress of the attack, while the latter counts the number of intercepted slices at any given moment. 

The figure shows that the attack progresses linearly: it starts by intercepting the first $k_1$ slices, each with probability $a_i$ given that the slice is sent to server/provider $i$. Having intercepted less than $k_1$ slices, there is no possibility to reconstruct the message body. After intercepting $k_1$ slices the next chain of states repeatedly alternate these steps: first it tries to reconstruct the message body with probability $x_j$, given $j = \var{ctr}_a$; if it fails it tries to intercept a new slice. The state $\var{pc}_a = \varval{done}$ denotes that the intruder reconstructed the message body.

\subsection{The provider attacker}

Let $\ServerAttacker$ be the MDP encoding the second type of intruder. The reason for its name is that it tries to obtain the credentials of the storage provider, and later it will intercept all the slices traveling towards that provider.

The intruder is depicted in Fig. \ref{fig:attacker2}. It has two local variables, viz. $\var{pc}_a$ and $\var{ctr}_a$. Similarly to the previous intruder, the former variable models the progress of the attack, while the latter counts the number of intercepted slices.

The intruder has an initial chain of $m$ states where it tries to attack every provider tossing a coin with probability $a_i$; if the attack is successful it sets a flag $\var{att}_a^i$ for provider $i$. State $\var{pc}_a = m$ is reached when all attack attempts are decided (some succeeded and some failed). In a loop the attacker synchronizes with the $busy$ action from the client that is sending a message, to intercept every slice sent to an attacked provider. From state $(\var{pc}_a = m, \var{ctr}_a = j)$ there is a transition to some state with $\var{pc} = \varval{done}$ labeled with probability $x_j$, meaning that it has probability $x_j$ to reconstruct the content of the message, given $j$ intercepted slices.

\begin{figure*}[tb]
\centering
\resizebox{0.8\linewidth}{!}{\begin{tikzpicture}[->,>=stealth',shorten >=1pt,auto,node distance=4cm,
                    semithick]
  \tikzstyle{every state}=[circle, minimum width=30pt, fill=white,draw=black,text=black]
  \tikzstyle{ghost}=[circle, minimum width=30pt, fill=white,draw=white,text=black]

  \node[state]  (A)                    {$0$};
  \node[state]  (B) [right of=A]       {$1$};
  \node[ghost]  (C) [right of=B]       {$\cdots$};
  \node[state]  (D) [right of=C]       {$m$};
  \node[state]  (E) [right of=D]       {$\var{done}$};
  \node[ghost]  (F) [left=1.5cm of A]  {};

  \tikzstyle{lblabove}=[align=center,midway,above]
  \tikzstyle{lblbelow}=[align=center,midway,below]

  \path (A) edge                node [lblbelow] {$\{ 1-a_1 \}$}
                                                                (B)
            edge [bend left]    node [lblabove] {$\{ a_1 \} \var{att}_a^1 = 1$}
                                                                (B)
        (B) edge                node [lblbelow] {$\{ 1-a_2 \}$}                        (C)
            edge [bend left]    node [lblabove] {$\{ a_2 \} \var{att}_a^2 = 1$}
                                                                (C)
        (C) edge                node [lblbelow] {$\{ 1-a_m \}$}
                                                                (D)
            edge [bend left]    node [lblabove] {$\{ a_m \} \var{att}_a^m = 1$}
                                                                (D)
        (D) edge [loop below]   node [lblbelow] {$[busy]$\\$\exists i \in [1,m]. \var{s}_c = i \wedge \var{att}_a^i = 1 \wedge \var{ctr}_a' = \var{ctr}_a + 1$}
                                                                (D)
            edge [loop above]   node [lblabove] {$[busy]$\\$\exists i \in [1,m]. \var{s}_c = i \wedge \var{att}_a^i = 0$}
                                                                (D)
            edge                node [lblabove] {$\{ x_{\var{ctr}_a} \}$\\$\var{ctr}_a \ge k_1$}
                                                                (E)
        (F) edge                node [lblabove] {$\var{att}_a^1=0$\\$\ldots$\\$\var{att}_a^m=0$}
                                                                (A);
				
\end{tikzpicture}}
\caption{\label{fig:attacker2}The MDP $\ServerAttacker$}
\end{figure*}

\section{Parametric formal verification}
\label{sec:formal}

In Section \ref{sec:modeling} we have seen that the problem at our hands is intrinsically parametric. The model checking problem requires its input MDP to be finite, thus we must fix the system parameters. On the other hand, this means that the outcome of our formal verification holds only for the specific configuration of the parameters themselves. One of the common desiderata when doing parametric formal verification, is to prove \emph{universal properties}, i.e. we should check whether some property holds for any configuration of parameters. In this work we are able to measure the confidentiality of dispersal algorithms for any number of storage providers in presence of a slice attacker, while in the case of the provider attacker the degree of confidentiality depends on the actual number of storage providers in the network.
Before showing the detailed formal analysis, we report the needed formal ingredients.

\subsection{Preliminaries}

$\PCTLstar$ is a temporal logic for describing qualitative and quantitative aspects of probabilistic systems.
The grammar of $\PCTLstar$ formulae is the following:
\[
\begin{array}{ccl}
\Phi &::=& \true \ |\ p \ |\ \Phi \wedge \Phi \ |\ \neg \Phi \ |\ \Prob_J(\varphi) \\ 
\varphi &::=& \Phi \ |\ \varphi \wedge \varphi \ |\ \neg \varphi \ |\ \ctlnext \varphi \ |\ \ctlglob \varphi \ |\ \ctlfuture \varphi \\
\end{array}
\]
where $p \in \AP$ and $J \subseteq [0,1]$ is a rational interval. Terms of $\Phi$ are \emph{state formulae}, while terms of $\varphi$ are \emph{path formulae}.

A thorough description of the logic satisfiability relation is beyond the aims of this paper, since the subject is covered by several textbooks (e.g. see \cite[Ch. 10.4]{BaierKatoen}). 
Intuitively, formula $\ctlglob \varphi$ holds w.r.t. some path iff every state visited in the path satisfies the sub-formula $\varphi$. Formula $\ctlfuture \varphi$, instead, holds w.r.t. some path iff some visited state satisfies sub-formula $\varphi$. The state formula $\Prob_{[a,b]}(\varphi)$ holds w.r.t. state $s$ iff the sub-formula $\varphi$ holds in all paths starting from $s$ with some probability $p \in [a,b]$.
%
%
Given an MDP $\proc$, let us write $\proc \models \Phi$ to express that \emph{all the initial states} of $\proc$ satisfy the property $\Phi$.

Given a $\PCTLstar$ \emph{path} formula $\varphi$ and an MDP $\proc$, there exist polynomial time algorithms computing the minimum and maximum probabilities of $\varphi$ w.r.t. all the initial states of $\proc$ \cite[Ch. 10.6]{BaierKatoen}. In the following we will write $\Probmin(\varphi,\proc)$ and $\Probmax(\varphi,\proc)$ to denote such computed probabilities. \footnote{Note that, in general, given any $\PCTLstar$ formula $\varphi$ and MDP $\proc$, it is possible that $\Probmin(\varphi, \proc) \neq \Probmax(\varphi, \proc)$. This is a consequence of the sequence of non-deterministic choices that can be taken in the executions of $\proc$, each leading to a (possibly) different probability outcome associated to $\varphi$. This motivates the interest in discovering the minimum and maximum probabilities with which $\varphi$ holds in $\proc$.}

From the definitions of $\proc \models \Phi$, $\Probmin$ and $\Probmax$, the following fact holds immediately.

\begin{proposition}[{\cite[Ch. 10.6]{BaierKatoen}}]
\label{prop:mc_vs_minmax}Given any MDP $\proc$, any $\PCTLstar$ path formula $\varphi$, and any $0 \le a \le b \le 1$, then:
\[
\proc \models \Prob_{[a,b]} \varphi  \iff a \le \Probmin(\varphi,\proc) \wedge b \ge \Probmax(\varphi,\proc)
\]
\end{proposition}

\begin{example}
\label{ex:spec}
Assume two MDPs $\proc_1 := \Client \comp \SliceAttacker$ and $\proc_2 := \Client \comp \ServerAttacker$. Assume a proposition $\prop{hacked} \in \AP$ and labeling $L_1$ of $\proc_1$ (resp. $L_2$ of $\proc_2$) such that $\prop{hacked} \in L_1(s)$ (resp. $\prop{hacked} \in L_2(s)$) iff $s.\var{pc}_a = \varval{done}$.  We can measure the likelihood of breaking the confidentiality requirement of $\proc_1$ (resp. $\proc_2$) computing  $\Probmin(F(\prop{hacked}), \proc_1)$ and $\Probmax(F(\prop{hacked}), \proc_1)$ (resp. $\Probmin(F(\prop{hacked}), \proc_2)$ and $\Probmax(F(\prop{hacked}), \allowbreak \proc_2)$).
\end{example}

Due to Proposition \ref{prop:mc_vs_minmax}, the \emph{probabilistic model checking} problem may assume two different flavours:
\begin{itemize}
\item \emph{qualitative}: take as input an MDP $\proc$ and a formula $\Phi$, and return $\true$ iff $\proc \models \Phi$;
\item \emph{quantitative}: take as input an MDP $\proc$ and a path formula $\varphi$, and compute $\Probmin$ and $\Probmax$.
\end{itemize}
Here we use the \emph{quantitative} probabilistic model checking.

Given an MDP $\proc$, one can show that two states $s$ and $s'$ are indistinguishable, from a probabilistic point of view, if (i) every step taken from $s$ is mimicked by some step taken from $s'$, (ii) both steps end in equivalent states, and (iii) the viceversa is also true. This is captured by the notion of \emph{probabilistic bisimulation}.

\begin{definition}[{\bf Probabilistic Bisimulation}, {\cite[Ex. 10.27]{BaierKatoen}}]
\label{def:pbisim}
Given an MDP $\MDP$, a probabilistic bisimulation is an \emph{equivalence relation} $R \subseteq S \times S$ such that, for any $s,s' \in S$, $R(s,s')$ iff:
\begin{itemize}
\item $L(s) = L(s')$, and 
\item $\MDPprob(s,\act,X) = \MDPprob(s',\act,X)$, $\forall \act \in \MDPact, X \in \quotient{S}{R}$
\end{itemize}
where $\quotient{S}{R}$ represents the \emph{quotient set} of $S$ by $R$.
\end{definition}

Given two states $s,t$, let us write $s \bsim_R t$ if $R(s,t)$ for some probabilistic bisimulation $R$. When $R$ is clear from the context, we may omit it. The given definition of probabilistic bisimulation helps establishing that two states of the same MDP cannot be distinguished. It is possible to use probabilistic bisimulation to check whether different MDPs are indistinguishable. Given two MDPs, $\proc_1 = \MDPi{1}$ and $\proc_2 = \MDPi{2}$, let $S := S_1 \uplus S_2$ be the disjoint union of the state sets. Let us write $\proc_1 \bsim_R \proc_2$ iff $\AP_1 = \AP_2$ and there exists a bisimulation relation $R \subseteq S \times S$ and $\iota_1(X) = \iota_2(X)$ for each $X \in \quotient{S}{R}$.
It is known that bisimilar MDPs satisfy the same $\PCTLstar$ formulae.

\begin{theorem}[{\cite[Ex. 10.27]{BaierKatoen}}]
\label{th:bisim}
Given two MDPs $\proc_1$ and $\proc_2$ such that $\proc_1 \bsim \proc_2$, then $\proc_1 \models \Phi$ iff $\proc_2 \models \Phi$, for any $\Phi \in \PCTLstar$.
\end{theorem}

Proposition \ref{prop:mc_vs_minmax} and Theorem \ref{th:bisim} yield the following.

\begin{restatable}{corollary}{BisimMinMax}
\label{cor:bisim_preserves_minmax}
Given two MDPs $\proc_1$,$\proc_2$ such that $\proc_1 \bsim \proc_2$ and any $\PCTLstar$ path formula $\varphi$, then $\Probmin(\varphi, \proc_1) = \Probmin(\varphi, \proc_2)$ and $\Probmax(\varphi, \proc_1) = \Probmax(\varphi, \proc_2)$.
\end{restatable}

\subsection{Parameter abstraction for the slice attacker}

Let us consider the MDP $\procslice = \Client \comp \SliceAttacker$.
In order to generalize our verification for any number of storage servers, we group the latter in \emph{channels}, i.e. collections of servers that are indistinguishable.

Formally we define a channel as a triple $(x, g, a)$ where $x \in \NatPos$ is the number of servers in the channel, $g : [1,x] \to [0,1]$ is a probability distribution and $a \in [0,1]$ is a probability value. Intuitively, $x$ is the number of storage servers belonging to the channel, $g$ is the probability distribution of picking any server in the channel when sending a slice, conditioned by the fact that the current channel has been chosen, and $a$ is the probability of attacking any server belonging to the channel. It is easy to see that if a channel has size one, it can only be defined as follows: $(1, \equiprob_1, a)$, for some $a \in [0,1]$.


In this analysis we assume that any storage server can host any number of slices. We also fix a subset of the model variables: $V = \{ \var{pc}_c, \var{ctr}_c, \var{pc}_a, \var{ctr}_a \}$. 

Given an Markov Decision Process $\proc$, let us write $\proc_f((x_1,g_1,a_1),\ldots,(x_k,g_k,a_k))$ denoting a copy of it where storage servers are grouped in the given $k$ channels.

Our claim is that if we check formulae that look only at variables in $V$ (and in particular that do not look at variables $\var{s}_c$ and $\var{ctr}_c^i$, for $i \in [1,m]$), then a system with one server per channel is indistinguishable from a system with an arbitrary number of servers per channel. This means that the number of channels in the system defines an upper limit, or \emph{cutoff}, to the size of the model to be verified.

\begin{restatable}{theorem}{ChannelCutoff}
\label{th:sc1_bisim}
Fix a positive number $k$ and any probability distribution $f : [0,k] \to [0,1]$. For all $n_1,\ldots,n_k \in \NatPos$, distribution probabilities $g_j : [0,n_j] \to [0,1]$ ($j \in [1,k]$), probability values $a_1,\ldots,a_k \in [0,1]$, let $\proc_1 = \procslice_f((1,\equiprob_1,a_1),\ldots,(1,\equiprob_1,a_k))$ and $\proc_2 = \procslice_f((n_1,g_1,a_1),\ldots,(n_k,g_k,a_k))$. Let $\MDPlabel_1$ and $\MDPlabel_2$ be the respective labeling functions, and assume they are invariant with respect to $V$.
Then: $\proc_1 \bsim \proc_2$.
\end{restatable}

\ifextended
    For the sake of readability, the detailed proofs can be found in the appendix of this paper.
\fi

Next corollary follows from Theorem \ref{th:sc1_bisim} and Theorem \ref{th:bisim}.
\begin{corollary}
Given $\proc_1$ and $\proc_2$ as of Theorem \ref{th:sc1_bisim}, then for all $\Phi \in \PCTLstar$:
\[
\proc_1 \models \Phi \iff \proc_2 \models \Phi 
\]
\end{corollary}


\subsection{Variable abstraction for the provider attacker}

Here we focus on the scenario in which the intruder can break one or more providers, thus accessing all the slices they store. Two applicative examples of this scenario are the cooperative storage systems, where each node joining the network can receive some of the slices, and the interconnected cloud, where a client may decide to split the message among several cloud storage providers. In the latter case we ignore the fact the cloud storage provider may further distribute the received slices among its own servers, and treat each such provider as a single server.

The model checking problem: $\Probmin(\varphi, \Client \comp \ServerAttacker)$, where $\varphi$ encodes our confidentiality requirements, remains a problem parameterized by the actual numbers of storage providers (similarly for $\Probmax$).

On the other side, one can easily see that if every provider can host any number of slices (i.e. $c \ge n$) we can produce an indistinguishable model that is much smaller w.r.t. the original one, by simply dropping the variables $W = \{ \var{ctr}_c^i : i \in [1,m] \}$ from the local state of $\Client$. This reduces the problem to a feasible one, allowing us to experimentally measure the degree of confidentiality of the considered dispersal algorithm w.r.t. the number of slices $n$ and the number of providers $m$. Call $V$ the set of remaining variables (i.e. $V \cap W = \emptyset$).

Name $\Client'$ a copy of $\Client$ whose state does not contain variables in $W$. Intuitively, this means that $\Client'$ does not check whether a provider reached its capacity, but this is not a limitation since we assumed that every storage provider can host any number of slices. 

\begin{restatable}{theorem}{ServerAbstraction}
\label{th:sc2_bisim}
Assume MDPs $\proc_1 = \Client \comp \ServerAttacker$ and $\proc_2 = \Client' \comp \ServerAttacker$. Let $\MDPlabel_1$ and $\MDPlabel_2$ be the respective labeling functions, and assume they are invariant with respect to $V$. Then: $\proc_1 \bsim \proc_2$.
\end{restatable}


\begin{corollary}
Given $\proc_1$ and $\proc_2$ as of Theorem \ref{th:sc2_bisim}, then for all $\Phi \in \PCTLstar$: 
\[
\proc_1 \models \Phi \iff \proc_2 \models \Phi
\]
\end{corollary}


\section{Experiments}
\label{sec:experiments}

Here we show how to use model checking for measuring the likelihood of breaking the confidentiality requirement against systems of growing sizes. In particular the parameter $n$ is increased among runs. We remark that any cloud system is characterized by its own sets of parameters (e.g. the probabilities of attacking the used providers may be part of their SLA). In our experiments we choose parameters arbitrarily, mainly for showing the feasibility of the approach and underlining the weakness of data dispersal algorithms in some cloud environment , viz. interconnected clouds.\fs{This paragraph takes care of observation from Reviewer 1}
The checked specifications are taken from Example \ref{ex:spec}.

Thanks to the parametric abstractions explained in Section \ref{sec:formal}, the verification outcomes using the intruder $\SliceAttacker$ hold no matter the number of storage servers/providers in the system. In the case of intruder $\ServerAttacker$ the results depend on the number of storage providers in the network.

The experiments were run on a machine Xeon Quad Core 2.3 Ghz with 8 GB RAM and Linux 2.6.32 64 bit. The points in the graphs correspond to distinct instances of the model checking problem, all requiring from few seconds to 60 minutes to complete.

\begin{figure*}
\begin{minipage}{0.5\linewidth}
\centering
\includegraphics[width=\linewidth]{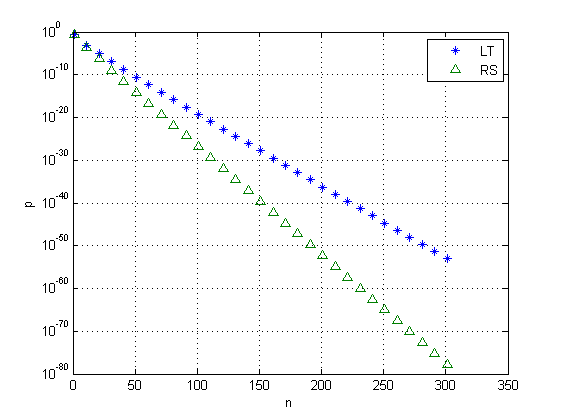}
\caption{\label{fig:rs_vs_lt} Compare AONT-RS vs. AONT-LT using $\SliceAttacker$}
\end{minipage}
\begin{minipage}{0.5\linewidth}
\centering
\includegraphics[width=\linewidth]{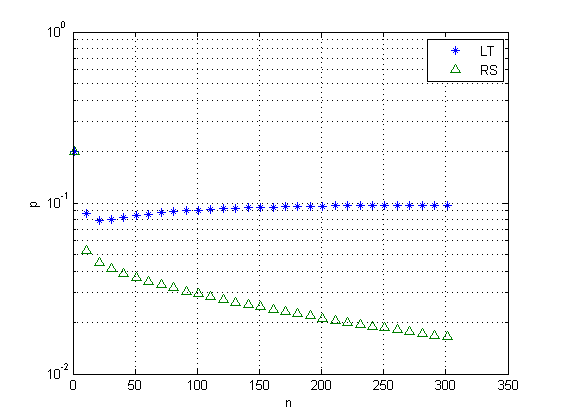}
\caption{\label{fig:rs_vs_lt_cloud} Compare AONT-RS vs. AONT-LT using $\ServerAttacker$}
\end{minipage}
\end{figure*}

In the first set of experiments we compare two different types of dispersal algorithms, those based on RS transforms against those based on LT transforms. For this analysis we fixed the number of channels/providers to $3$ and the attacking probabilities to $a_1 = 0.1$, $a_2 = 0.2$, and $a_3 = 0.3$. In the RS case we assumed $k_1 = k_2 = 0.7 \cdot n$, while in the LT case we assumed $k_1 = 0.6 \cdot n$ and $k_2 = 0.8 \cdot n$. In the case of the LT transforms we assumed that the sequence of probabilities $x_j$, for $j \in [k_1,n]$ is defined as follows: let $\delta = \frac{1}{k_2 - k_1 + 1}$, $x_j = (\min(j,k_2) - k_1 + 1) \cdot \delta$. Intuitively, probabilities $x_j$ grow linearly in the interval $[k_1,k_2-1]$ and then stabilize at $1$ for values greater than or equal to $k_2$ .

In Fig. \ref{fig:rs_vs_lt} and \ref{fig:rs_vs_lt_cloud} we depict how the probability of breaking the confidentiality requirement varies w.r.t. $n$. Let us remark that $\Probmax$ and $\Probmin$ versions of the formula coincide in every point of the series. We also notice that the algorithm (under the given parameters) shows a high degree of confidentiality against $\SliceAttacker$, while it is sensibly less confidential against $\ServerAttacker$. Moreover, the confidentiality in the latter case, after some initial change, stabilizes and does not depend on the actual number of exchanged slices. The latter observation is not surprising since the intruder mainly attacks providers, and few providers will receive many slices, thus giving the intruder a high probability of guessing the message body.

\begin{figure*}
\begin{minipage}{0.5\linewidth}
\centering
\includegraphics[width=\linewidth]{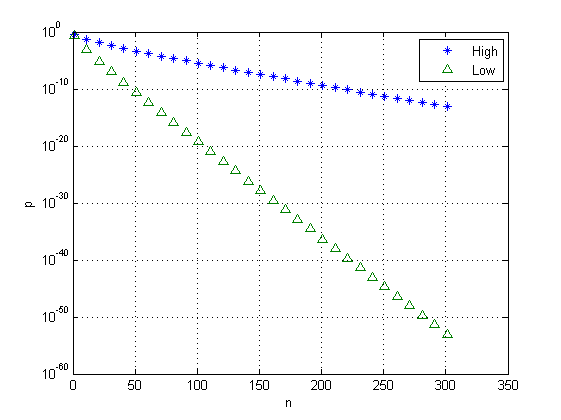}
\caption{\label{fig:prob_slice}Compare channel probability of attack using $\SliceAttacker$}
\end{minipage}
\begin{minipage}{0.5\linewidth}
\centering
\includegraphics[width=\linewidth]{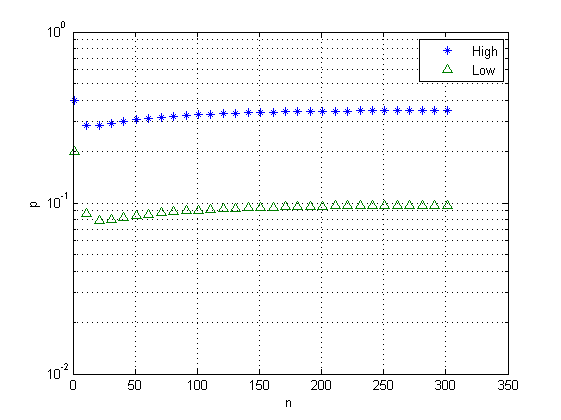}
\caption{\label{fig:prob_server}Compare provider probability of attack using $\ServerAttacker$}
\end{minipage}
\end{figure*}

The second block of experimental data compares the effect of different attack probabilities, viz. $a_1 = 0.1$, $a_2 = 0.2$, and $a_3 = 0.3$ (see Low) vs. $a_1 = 0.3$, $a_2 = 0.4$, and $a_3 = 0.5$ (see High).\fs{check names in figures are consistent} For this analysis we used only LT transforms and we fixed the number of channels/providers to $3$. As before, $k_1 = 0.6 \cdot n$ and $k_2 = 0.8 \cdot n$ and the series $x_j$ grows linearly as for the previous experiment. Fig. \ref{fig:prob_slice} and \ref{fig:prob_server} summarize the model checking outcomes. As would be expected, higher probabilities of intercepting slices give the data dispersal algorithm a very low level of confidentiality against intruder $\SliceAttacker$. Again, the low number of providers causes a low degree of confidentiality against intruder $\ServerAttacker$.

\begin{figure*}
\begin{minipage}{0.5\linewidth}
\centering
\includegraphics[width=\linewidth]{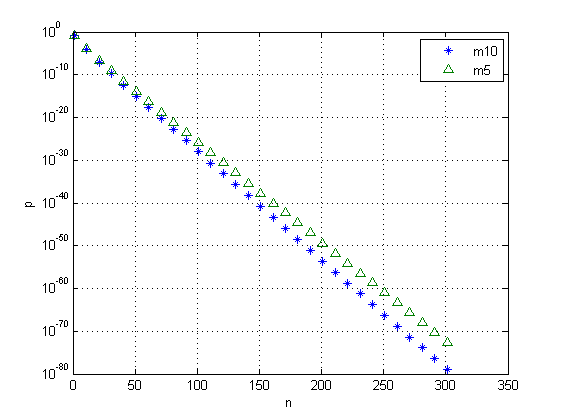}
\caption{\label{fig:chann_slice}Compare different number of channels using $\SliceAttacker$}
\end{minipage}
\begin{minipage}{0.5\linewidth}
\centering
\includegraphics[width=\linewidth]{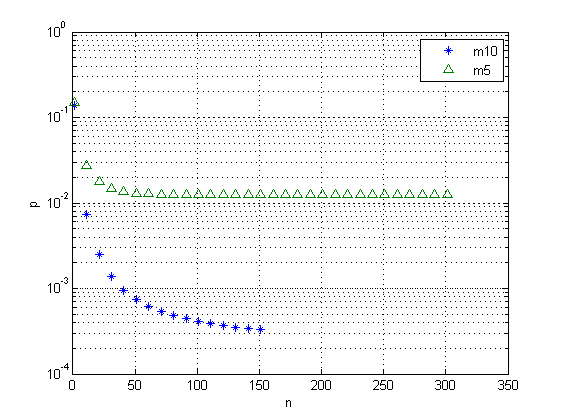}
\caption{\label{fig:chann_server}Compare different number of providers using $\ServerAttacker$}
\end{minipage}
\end{figure*}

Finally, a third set of experiments compares the effects of different numbers of channels/providers in the system and the results are given in Fig. \ref{fig:chann_slice} and \ref{fig:chann_server}. In both cases a LT transform was used, with $k_1 = 0.6 \cdot n$ and $k_2 = 0.8 \cdot n$. In one case we assumed 5 channels/providers in the network (see m5) and in the other 10 channels/providers (see m10). The probabilities $a_i$, for $i \in [1,m]$, are distributed uniformly in the interval $[0,0.25]$ in both cases. $x_j$ is defined as for the previous experiment. Once again the degree of confidentiality against $\ServerAttacker$ is considerably lower than that against $\SliceAttacker$. We underline that the experiment with 10 providers could be verified only for a small number of slices ($n \le 150$), before running out of memory. Even though this fact represents a scalability issue of the presented methodology, the verification outcomes are still of practical interest since interconnected (or federated) cloud solutions usually employ a limited number of storage providers. Confidentiality of data dispersal with a small number of storage nodes appears to be too weak against attacks directed to the storage provider.

\section{Related work}
\label{sec:rw}

Formal verification of security requirements has a long history. 
In this area, model checking plays a predominant role
\cite{Basin2011,Meadows2006,Pagliarecci2013,Panti2002}.
The traditional approaches consist in model checking the security requirements 
of a system opposed to an adversary able to intercept, remove, modify the original messages as well as to inject new messages.
In this respect, the Dolev-Yao intruder model \cite{Dolev1983} is considered the most general model (the worst case) \cite{Cervesato2001}
as it assumes a non-deterministic attacker in full control of the communication channels.

Traditional model checking, though, is not suitable for verifying security of cloud systems: it can only verify whether a system can be attacked or not. We \emph{assume}, instead, that every component of a cloud system can be attacked with some degree of probability, and are more interested in measuring the likelihood of such attacks. This motivated us to define custom probabilistic intruder models, in place of the Dolev-Yao intruder.
To the best of our knowledge, few authors used probabilistic model checking for measuring security of systems \cite{Shmatikov2004,Lenzini2015,Ouchani2015,Yang2016}.
\fs{this compares our work with traditional model checking, as required by Reviewer 3}

It is well known that model checking techniques must face the state-explosion problem, that easily makes the verification of real-world protocols and systems unfeasible. To overcome this limitation, one looks for \emph{abstraction techniques} \cite{Clarke1994} that reduce the description of the system to a feasible state-space, still preserving the relevant properties. Special forms of abstractions are required when the system state-space depends on given parameters and one wants to check whether some property holds for all values of such parameters \cite{Aminof2014,Spalazzi2014}. 

With regards to probabilistic models, several approaches use abstraction techniques. Legay et al. \cite{Legay2010} and Nouri et al. \cite{Nouri2014}, for example, collect traces of real or simulated systems. Next they sample them in order to build an MDP that abstracts the original system. Such technique avoids to build a complete analysis of all the traces for large-scale (or even infinite) systems. The larger is the sample, on the other side, the higher is the probability that the verification outcomes are correct.
Herd et al. \cite{Herd2015}, instead, proposed a trace sampling technique combined with trace fragmentation, i.e. only few fragments of a trace are considered.
Abate et al. \cite{Abate2014} proposed a method for transforming an MDP with an uncountable number of states into a Markov chain by means of a quotient-set based abstraction.
The paper proves that the produced Markov chain approximates a probabilistic bisimulation of the original MDP. \fs{check}

Finally, let us remark that also the structure of the attacker may determine the feasibility of the verification of security properties. In our work we employed a passive intruder model, and indeed several authors agree that this is enough when analyzing confidentiality requirements. For example, Li and Pang \cite{Li2015}, and Shmatikov \cite{Shmatikov2004} used passive intruders to verify anonymity of protocols, a special case of confidentiality. The latter work also considers probabilistic attacks. As far as we know, the use of a probabilistic passive attacker model for the analysis of data confidentiality 
is original.

\section{Conclusions}
\label{sec:conclusions}

We presented a unified framework for the probabilistic model checking of a broad class of data dispersal algorithms in interconnected or cooperative cloud storage systems.

We verified confidentiality requirements of dispersal algorithms, checking the likelihood that an intruder has of intercepting slices of information and reconstruct the information. 

In our framework we defined two types of probabilistic intruder, one tries to intercept each slice independently and the other attempts to attack the storage provider. In the former case the attack surface is the set of slices while in the latter it is the set of providers.

The problem is inherently parametric, since the CPU time and memory required to complete the verification are highly affected by several parameters, e.g. the number of slices used to split the information and the number of servers/providers.

By proving a probabilistic bisimulation property, we were able to generalize the results of the verification of confidentiality against the slice attacker to any number of servers in the network. The key observation, codified in our \emph{channel abstraction}, is that any group of servers sharing the same probability of being eavesdropped may form a channel and behave like a single server hosting all the slices.

The analysis of the confidentiality against the provider attacker suggests, on the contrary, that classic data dispersal algorithms may not be the best solution to ensure confidentiality in interconnected cloud environments, unless the number of storage providers is considerably high. 

We should remark that the conducted experiments fix some parameters to specific values. The conclusions thus are not fully generalizable w.r.t. such parameters. Among the modifiable parameters we remark that only $k_1$ and $k_2$ affect the state space and thus the complexity of the model checking problem. 

We leave as future research the investigation of better abstractions suitable for the verification of the confidentiality of data dispersal algorithms against $\ServerAttacker$ on networks with many storage providers. That would improve the scalability of our framework to handle the case of cooperative storage systems.

\bibliographystyle{IEEEtran}
\bibliography{IEEEabrv,lit}

\ifextended
\newcolumn
\appendix

\section{Proofs of Theorems}

\BisimMinMax*
\begin{IEEEproof}
Assume MDPs $\proc_1$ and $\proc_2$ such that $\proc_1 \bsim \proc_2$, and any path formula $\varphi \in \PCTLstar$. Let us name $a = \Probmin(\varphi, \proc_1)$ and $a' = \Probmin(\varphi, \proc_2)$. Assume, by contradiction and w.l.o.g., that $a > a' \ge 0$.

It is immediate that some $a \le b \le 1$ must exist such that: $\proc_1 \models \Prob_{[a,b]} (\varphi)$, thus $\proc_2 \models \Prob_{[a,b]}(\varphi)$ (by Theorem \ref{th:bisim}).

As we observed earlier (see Proposition \pageref{prop:mc_vs_minmax}), by definition of the model checking problem $\PCTLstar$, $\Probmin$ and $\Probmax$ (see e.g. \cite{BaierKatoen}), the following holds: $\proc \models \Prob_{[x,y]}(\varphi)$ iff $x \le \Probmin(\varphi, \proc)$ and $y \ge \Probmax(\varphi, \proc)$, for any MDP $\proc$, $\PCTLstar$ path formula $\varphi$ and rationals $0 \le x,y \le 1$.

In our case, and from $\proc_2 \models \Prob_{[a,b]}(\varphi)$, it follows that:
\[
a \le \Probmin(\varphi, \proc_2) = a' 
\]
which in turn contradicts our assumption $a > a'$. This proves that $\Probmin(\varphi, \proc_1) = \Probmin(\varphi, \proc_2)$.

Using a symmetric argument, it is straightforward to prove that, given the corollary assumptions, $\Probmax(\varphi, \proc_1) = \Probmax(\varphi, \proc_2)$.

\end{IEEEproof}

\ChannelCutoff*

\begin{IEEEproof}
Let $n := \sum_{i=1}^{k} n_i$. Let $H : [1,n] \to [1,k]$ be a surjective mapping. Intuitively, $H(j) = i$ iff $j$ belongs to the $i$-th channel. If for some $j,k$, $H(j) = H(k)$ it means that $(p_j,q_j) = (p_k,q_k)$, i.e. they belong to the same channel, and we will write $j =_H k$.
In this proof we call \emph{transition template} a PRISM transition, possibly with indexed variables. Let us call $\Theta$ the set of the above transition templates. Given any $\theta \in \Theta$ and a state $s$, we write $\theta(s)$ to denote the actual MDP transition obtained by instantiating the variables in $\theta$ with their values in state $s$. It is evident that for any $\theta$ and $s$ there exists a unique target state $s'$. For example: let $\theta$ be an indexed PRISM transition $[\act] \var{foo}_i=1 \to p : \var{foo}'_i=2$, and an MDP state $s$ such that $s.\var{foo}_1=1$ and $s.\var{foo}_2=1$, then $\theta(s)$ is the MDP transition: $(s,p,\act,s')$ such that $s'.\var{foo}_1 = 2$, and $s'.\var{foo}_2 = 2$. \fs{check: a transition template may not be applicable to a state (e.g. if the lhs does not match)}

Given a transition $\tau = (s,\alpha,p,s') \in \MDPprob$ we will denote with $\pre(\tau)$ the source state $s$ and with $\post(\tau)$ the target state $s'$.

Name $\MDPstates_1$ the set of states of $\proc_1$, $\MDPstates_2$ the set of states of $\proc_2$, and $\MDPstates$ their disjoint union. Let us define the relation $R \subseteq \MDPstates \times \MDPstates$ as follows: $R(s_1,s_2)$ iff the following properties hold:
\begin{itemize}
\item $s_1 \equiv_V s_2$ 
\item $s_1.\var{s}_c =_H s_2.\var{s}_c$
\item $\forall i \in [1,m].\ s_1.\var{ctr}_i^1 = \sum_{j=1}^{n_i} s_2.\var{ctr}_i^j$
\end{itemize}

In order to show that $R$ is a probabilistic bisimulation, let us first observe that $R$ is an equivalence relation (i.e. it is transitive, reflexive and symmetric).

Secondly, let us underline that $L(s_1) = L(s_2)$ follows by our definition of $\AP$ and the fact that $\forall v \in V.\ s_1.v = s_2.v$.

Finally, let us show that: $\MDPprob(s_1,\act,X) = \MDPprob(s_2,\act,X)$, for all $X \in \quotient{S}{R}$ and all $\act \in \MDPact$.

Let us consider two cases: either (C1) $X = [s_1]_R$, or (C2) $X \neq [s_1]_R$. Notice that by definition of $X$ and $R$, $[s_1]_R = [s_2]_R$.

Case (C1) is possible iff there is a transition $\tau$ satisfying all the following conditions:
\begin{itemize}
\item $\pre(\tau) \equiv_V \post(\tau)$ i.e. $\tau$ does not change any variable $v \in V$ (otherwise $s_1,s_2 \not\in X$);
\item $\forall i \in [1,n].\ \pre(\tau) \equiv_{\{ \var{ctr}_i \}} \post(\tau)$, i.e. $\tau$ does not change any counter (otherwise the counters can only increase, thus their sum increase, and thus $s_1,s_2 \not\in X$);
\item either $\pre(\tau) = \post(\tau)$ or $\pre(\tau) \not\equiv_{\{\var{s}_c\}} \post(\tau)$ i.e. it is either a self-loop or it changes variable $\var{s}_c$.
\end{itemize}
Inspecting all the transitions in the model, there is none that satisfies the above conditions, thus this case is \emph{impossible}.

Let us split case (C2) in subcases. Either (C2.1) $X$ differs from $[s_1]$ only for variables in $V$, or (C2.2) $X$ differs from $[s_1]$ for variables in $V \cup \{ \var{ctr}_i, \var{s}_c \}$.

Case (C2.1) is possible for the following transition templates:
\begin{itemize}
\item any transition of $\SliceAttacker$, or
\item the $\Client$ transition template: $(\var{pc}_c = 0, \var{ctr}_c = \varval{n}) \to (\var{pc}_c'=3, \var{s}_c=0)$
\end{itemize} 


Let us observe the following facts, for any $\theta \in \Theta$:
\begin{itemize}
\item $\exists q_1 \in X, p \in [0,1].\ \theta(s_1) = (s_1,p,\act,q_1) \Rightarrow \exists q_2 \in X.\ \theta(s_2) = (s_2,p,\act,q_2)$;
\item $\exists q_2 \in X, p \in [0,1].\ \theta(s_2) = (s_2,p,\act,q_2) \Rightarrow \exists q_1 \in X.\ \theta(s_1) = (s_1,p,\act,q_1)$;
\end{itemize}
This means, intuitively, that whenever $\tau$ is applicable to state $s_1$ and reaching some state $q_1 \in X$, it is also applicable to the equivalent state (by R) $s_2$ and reaches some state $q_2$ equivalent to $q_1$. This can be proven checking all transition templates in $\Theta$: their final statuses $q_1,q_2$ have the same values for variables in $V$ and they didn't change the values of variables $\{ \var{s}_c, \var{ctr}_i \}$ w.r.t. $s_1,s_2$. Since the correspondence preserves the value of probability $p$, it follows that:
\[
\begin{array}{lll}
\MDPprob(s_1,\act,X) &= \sum_{\stackrel{q \in X, \theta \in \Theta,}{\theta(s_1) = (s_1,p, q)}} p \\
&= \sum_{\stackrel{q' \in X, \theta' \in \Theta,}{\theta'(s_2) = (s_2,p,q')}} p &= \MDPprob(s_2,\act,X)
\end{array}
\]

Case (C2.2): X differs from $[s_1]$ for variables in $V$ and for variables in $\{ \var{s}_c, \var{ctr}_c^i \}$. The possible transition templates in this case are all $\Client$ templates: 
\begin{itemize}
\item $[\var{busy}] (\var{pc}_c=1, \var{s}_c=i, \var{ctr}_c^i < c) \to (\var{pc}'_c=0, \var{ctr}'_c=\var{ctr}_c+1,\var{s}'_c=0)$
\item $(\var{pc}_c=0, \var{ctr}_c < n) \to p_i : (\var{pc}'_c=1, \var{s}_c=i)$
\end{itemize}
In the case of the first template again we can directly check that:
\begin{itemize}
\item $\exists q_1 \in X, p \in [0,1].\ \theta(s_1) = (s_1,p,\act,q_1) \Rightarrow \exists q_2 \in X.\ \theta(s_2) = (s_2,p,\act,q_2)$;
\item $\exists q_2 \in X, p \in [0,1].\ \theta(s_2) = (s_2,p,\act,q_2) \Rightarrow \exists q_1 \in X.\ \theta(s_1) = (s_1,p,\act,q_1)$;
\end{itemize}
Take all transitions induced by the first template: the final statuses satisfy the property $q_1 \equiv_W s_1$ and $q_2 \equiv_W s_2$, where $W = V \cup \{ \var{s}_c \}$. When variable $\var{ctr}^j_c$ changes, for some $j$, it increases by one. In this case the same template $\theta$ can be applied to state $s_2$ and variable $\var{ctr}^h_c$ can increase its value by one (with same probability) for some other channel $h$, such that $h =_H j$.

In the case $\theta$ is the second template, let us call $n = \sum_{i=1}^{k}n_i$, i.e. $n$ represents the total number of servers belonging to some channel in the system. Let us define the set $A_s = \{ \tau : \tau \in \theta, \exists q.\ \MDPprob(s,\act,q) > 0 \}$, i.e. $A_s$ contains all the transitions that are enabled in $s$ and are induced by template $\theta$. Let us define the sets: $A_s^i = \{ \tau : \tau \in \theta, \tau = (s,p,\act,q), p > 0, \left< \var{ctr}^j_c < \var{c} \right> \in \tguard(\tau), j =_H i \}$. We have that:
\[
A_s = \bigcup_{i=1}^n A_s^i
\]
Being in this case, $s_1$ and $s_2$ are the states where transition of template $\theta$ is enabled and picks a channel to send the slice to\fs{formalise this statement}. We can observe that: $|A_{s_1}^i| = 1$ and $|A_{s_2}^i | = n_i$, for any $i \in [1,k]$. Intuitively: by definition, in the small system we have only one possible concrete transition that picks a server from channel $i$, while in the big system we have $n_i$ concrete transitions each choosing a different server from channel $i$. Thus, we can write that: $A_{s_1}^i = \{ (s_1,p,\act,q) \}$, for some $p \in [0,1]$ and $q \in S_1$, and $A_{s_2}^i = \{ (s_1,p_1,\act,q_1), \ldots, (s_1,p_{n_i},\act,q_{n_i}) \}$, for some $p_1, \ldots, p_{n_i} \in [0,1]$ s.t. $\sum_{j=1}^{n_i} p_j = p$, and $q_1, \ldots, q_{n_i} \in S_2$.

Finally, we can show that, for all $i$:
\[
\begin{array}{lll}
\MDPprob(s_1,\act,X) &= \MDPprob(s_1,\act,q) = p \\
    &= \sum_{j=1}^{n_i} \MDPprob(s_2,\act,q_j) \\
    &= \sum_{j=1}^{n_i} p_j &= \MDPprob(s_2,\act,X)
\end{array}
\]
In fact: $\sum_{j=1}^{n_i} p_j = \sum_{j} g_i(j) \cdot f(i) = f(i) \cdot \sum_j g_i(j) = f(i)$, since $1 = \sum_j g_i(j)$, by our assumptions.
\end{IEEEproof}

\ServerAbstraction*

\begin{IEEEproof}
First of all, let us observe that variables $\var{ctr}_c^i$ is always compared with parameter $c$, the storage capacity. Since by assumption $c > n$, for all $n \in \NatPos$, this guard can always be dropped from $\Client$ as it is a tautology. What remains is a MDP that updates the variables $\var{ctr}_c^i$ but never reads it.

The MDP $\Client'$ is identical to \Client, except that it does not update the variables $\var{ctr}_c^i$. It is immediate to see that every transition enabled in $\Client$ must be enabled also in $\Client'$, and viceversa.

Name $\MDPstates_1$ the set of states of $\proc_1$, and $\MDPstates_2$ the set of states of $\proc_2$, and name $\MDPstates = \MDPstates_1 \uplus \MDPstates_2$ their disjoint union. Let us define the relation $R \subseteq \MDPstates \times \MDPstates$ as follows: $R = \{ (s_1,s_2) : s_1 \equiv_V s_2 \}$.

By our assumptions, $\MDPlabel_1$ and $\MDPlabel_2$ are invariant w.r.t. variables in $V$, meaning that $s_1 \equiv_V s_2 \Rightarrow \MDPlabel_1(s_1) = \MDPlabel_2(s_2)$. Combined with definition of $R$ we have that $R(s_1,s_2) \Rightarrow \MDPlabel_1(s_1) = \MDPlabel_2(s_2)$.

Since every transition enabled in $\proc_1$ is also enabled in $\proc_2$, and since they have the same probability, the second requirement of a probabilistic bisimulation holds (see Definition \ref{def:pbisim}).

\end{IEEEproof}

\fi

\end{document}